\documentclass[aip,jcp,amsmath,amssymb,preprint,longbibliography]{revtex4-1}
\usepackage{graphicx}
\usepackage[caption=false]{subfig}
\usepackage{notes2bib}
\newcommand\p{\mathbf{p}}
\newcommand\pbar{\mathbf{\bar{p}}}
\newcommand\dtheta{\chi_{\p\pbar}}

\newcommand\fourint{\int d\pbar\int d\p}
\newcommand\ddd{d_{\p}^{\pbar}}
\newcommand\aaa{a_{\p}^{\pbar}}

\begin{document}
\title{Necessary symmetry conditions for the rotation of light}
\author{Ivan Fernandez-Corbaton}
\email{ivan.fernandez-corbaton@mq.edu.au}
\affiliation{Department of Physics \& Astronomy, Macquarie University, NSW 2109, Australia}
\affiliation{ARC Center for Engineered Quantum Systems, NSW 2109, Australia}
\author{Xavier Vidal}
\affiliation{Department of Physics \& Astronomy, Macquarie University, NSW 2109, Australia}
\author{Nora Tischler}
\affiliation{Department of Physics \& Astronomy, Macquarie University, NSW 2109, Australia}
\affiliation{ARC Center for Engineered Quantum Systems, NSW 2109, Australia}
\author{Gabriel Molina-Terriza}
\affiliation{Department of Physics \& Astronomy, Macquarie University, NSW 2109, Australia}
\affiliation{ARC Center for Engineered Quantum Systems, NSW 2109, Australia}

\begin{abstract}
\noindent Two conditions on symmetries are identified as necessary for a linear scattering system to be able to rotate the linear polarisation of light: Lack of at least one mirror plane of symmetry and electromagnetic duality symmetry. Duality symmetry is equivalent to the conservation of the helicity of light in the same way that rotational symmetry is equivalent to the conservation of angular momentum. When the system is a solution of a single species of particles, the lack of at least one mirror plane of symmetry leads to the familiar requirement of chirality of the individual particle. With respect to helicity preservation, according to the analytical and numerical evidence presented in this paper, the solution preserves helicity if and only if the individual particle itself preserves helicity. However, only in the particular case of forward scattering the helicity preservation condition on the particle is relaxed: We show that the random orientation of the molecules endows the solution with an effective rotational symmetry; at its turn, this leads to helicity preservation in the forward scattering direction independently of any property of the particle. This is not the case for a general scattering direction. These results advance the current understanding of the phenomena of molecular optical activity and provide insight for the design of polarisation control devices at the nanoscale.
\end{abstract}
\maketitle
An object which cannot be superimposed onto its mirror image is said to be chiral. Chirality is entrenched in nature. For instance, some interactions among fundamental particles are not equivalent to their mirrored versions \cite{Wu1957}. Also, the DNA, and many aminoacids, proteins and sugars are chiral. The understanding and control of chirality has become important in many scientific disciplines. In chemistry, the control of the chiral phase (left or right) of the end product of a reaction is crucial, since the two versions can have very different properties. In nanoscience and nanotechnology, chirality plays an increasingly important role \cite{Noguez2011,Zhang2005}. 

Electromagnetic waves can also be chiral. The electromagnetic field has its chiral character mapped onto a binary property that can be related to the polarisation handedness of all the plane waves composing an electromagnetic field: Its helicity \cite{Tung1985,Weinberg1995}. From these considerations, and since electromagnetic waves are routinely used to interact with matter at the nano, meso, molecular and atomic scales, it is not surprising that the interaction between chiral light and chiral matter has become an important subject of study. Interestingly, the subject is quite old and, from the beginning, has always been associated with the rotation of the linear polarization of light. For instance, Biot discovered that when light propagates through a solution of certain types of molecules, its linear polarization rotates \cite{Biot1815}. Commonly referred to as molecular optical activity, the study of its root causes has a long history\cite{Ingold1967,Oloane1980,Barron2004}. In 1848, Pasteur identified the absence of mirror planes of symmetry of the molecule as a necessary condition \cite{Pasteur1848}. He called it ``dissym\'etrie mol\'eculaire'' and by it Pasteur meant non-superimposability of the molecule and its mirror image, in other words: Chirality. Nowadays, this necessary condition is assumed to also be sufficient, and the exceptions to the rule are explained by other means \cite{Oloane1980,Bishop1993}. A comprehensive theoretical study of optical activity based in symmetry principles can be found in \cite{Barron2004}, and the modern theoretical and computational methods for optical activity calculations are reviewed in \cite{Autschbach2012}. Current investigations of optical activity in metamaterials \cite{Baev2007,Decker2010,Zhao2012,Ren2012} are aimed at obtaining polarization manipulation devices for integrated nanophotonics. 

In this article we rigorously study the necessary symmetries that an otherwise general linear system must meet in order to rotate linear polarization states. We find two conditions, the lack of at least one mirror plane of symmetry and invariance under electromagnetic duality transformations to be necessary symmetry conditions for such system. Duality invariance is equivalent to the preservation of the helicity of light. This conservation law was first established in \cite{Calkin1965,Zwanziger1968}. Please refer to Refs.\cite{FerCor2012b,FerCor2012,Zambrana2013,FerCor2013} for examples of the use of helicity and duality in the study of light-matter interactions. 

We then consider a mixture of randomly oriented replicas of a single particle. The aim is to identify the restrictions imposed on the individual particle by each of the two necessary conditions that the mixture as a whole must meet. We find that the lack of at least one mirror plane of symmetry of the mixture translates into the condition of ``dissym\'etrie'', i.e. chirality, for the individual particle. With respect to duality symmetry, we find that the mixture is not a dual symmetric system unless the individual particle itself preserves helicity, which does not happen in general \cite{FerCor2012,FerCor2013}. Since chirality of the particle is accepted as the only necessary and sufficient condition for molecular optical activity of a solution, and helicity conservation does not have a recognized role, there seems to be a conflict between our results and the current understanding of molecular optical activity. This conflict is completely resolved: Due to the large number of randomly oriented particles, the mixture acquires an effective rotational symmetry, which is shown to lead to the conservation of helicity in the forward scattering direction independently of any property of the individual particle. Therefore, in the forward scattering direction, a solution of a chiral molecule can exhibit optical activity without the individual molecule having to preserve helicity. We will give analytical arguments and provide numerical simulations that prove that, in general, all other scattering directions break helicity conservation. The polarisation transformation in those directions is qualitatively different from the forward scattering case. The solution as a whole cannot be considered to have duality symmetry: For a plane wave decomposition, duality symmetry means helicity preservation for all input and output momenta. In an ordered system or in any non-forward scattering direction, explicit helicity preservation by the individual particle is needed for optical activity. This has direct implications for the design of materials with artificial optical activity by means of ordered arrays.

\begin{figure}[ht]
\begin{center}
\includegraphics[scale=0.5]{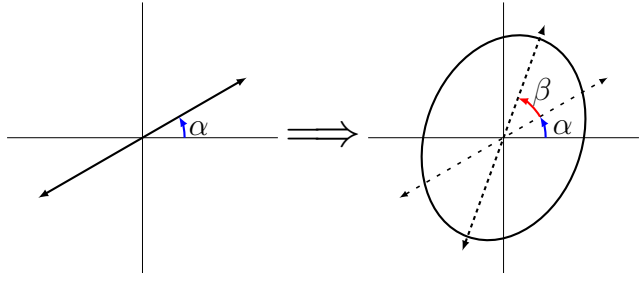}
\end{center}
\caption{(Color online) Transformation of input linear polarization states (diagram on the left) to rotated elliptical output polarization states (diagram on the right). The angle of rotation $\beta$ and the major to minor axis ratio (ellipticity) of the ellipse do not depend on the input angle $\alpha$.}
\label{fig:polrot}
\end{figure}

A clear definition of what we are referring to by the term ``optical activity'' is now in order. In this article, optical activity refers to the ability of a system to rotate the linear polarisation of light in a consistent manner: the incremental angle by which the input linear polarisation state is rotated at the output does not depend on the initial angle of the input linear polarisation. We also include in our definition of optical activity the possibility of circular dichroism by allowing the output polarisation to become elliptical, while the main axis of the ellipse still rotates in the aforementioned consistent manner. The output ellipticity is also independent of the input polarisation angle. This definition coincides for instance with the one given by Condon in its seminal work on optical rotation \cite{Condon1937}. What is excluded from the definition is, for example, rotation by an amount that depends on the initial angle of the input polarisation. Such transformations are also sometimes referred to as optical activity, for instance in some of the metamaterial literature \cite{Papakostas2003,Ren2012}. 

Throughout the paper, we will assume monochromatic electromagnetic fields with a harmonic $\exp(-i\omega t)$ time dependence. Consider the polarization transformation illustrated by Fig.\ref{fig:polrot}. A linearly polarized input transforms into an elliptically polarized output with its major ellipse axis rotated by a fixed angular quantity $\beta$ with respect to the angle $\alpha$ of the input linear polarization. Both the angle of rotation $\beta$ and the major to minor axis ratio (ellipticity) of the ellipse are independent of $\alpha$. Our aim is to identify what restrictions this class of transformations impose over a general conversion. This is most easily achieved expressing the input linear polarization state $[ \cos(\alpha)\, ,\, \sin(\alpha)]^{T}$ in the circular polarization basis 
\begin{equation}\nonumber
\begin{bmatrix} E_l\\E_r\end{bmatrix}=\frac{1}{\sqrt{2}}
\begin{bmatrix}
1 & i\\1&-i\\
\end{bmatrix}
\begin{bmatrix}\cos(\alpha) \\ \sin(\alpha)\end{bmatrix}=
\frac{1}{\sqrt{2}}\begin{bmatrix}\exp (i\alpha) \\ \exp (-i\alpha) \end{bmatrix},
\end{equation}
where a general transformation of the input reads
\begin{equation}
\label{eq:tx}
\begin{bmatrix}{F_l}\\{F_r}\end{bmatrix}
=
\begin{bmatrix}
a & b\\c&d\\
\end{bmatrix}
\begin{bmatrix} E_l\\E_r\end{bmatrix}
=\frac{1}{\sqrt{2}}
\begin{bmatrix}a\exp (i\alpha)+b\exp (-i\alpha)\\c\exp (i\alpha)+d\exp (-i\alpha) \end{bmatrix}.
\end{equation}

The angle of the major ellipse axis with respect to the horizontal axis is $\theta=\frac{1}{2}\arg{\left({F_l}{F_r}^*\right)}$. According to our specification of constant angle of rotation $2\theta=2(\alpha+\beta)$ for all $\alpha$, which then forces ${F_l}{F_r}^*=\eta \exp(i2(\alpha+\beta))$ where $\eta$ is a real number. Using (\ref{eq:tx}) we obtain the relationship:
{
\begin{equation}\nonumber
\label{eq:eler}
\begin{split}
	{F_l}{F_r}^*&=ac^*+ad^*\exp(i2\alpha)+bc^*\exp(-i2\alpha)+bd^*\\
	&=\eta \exp(i2(\alpha+\beta)),
\end{split}
\end{equation}
}
\noindent which must be valid for all $\alpha$ and hence imposes $b=c=0$ and gives $2\beta=2(\arg{a}-\arg{d})$. The most general matrix which meets the requirement is hence diagonal
\begin{equation*}
\begin{bmatrix}a & 0\\0&d\\\end{bmatrix}.
\end{equation*}
We conclude that our specified transformation is equivalent to the conservation of circular polarization states. From now on, we will refer to such a transformation as a generalized rotation of linear polarization, and abbreviate it by GRLP.

We now consider an electromagnetic scattering situation, where an incident field interacts with a linear scatterer. As a result of the interaction a scattered field is produced. Both incident and scattered fields can be decomposed in plane waves and can contain components in all directions. We now impose that the relationship between the polarisations of any pair of incident and scattered plane waves is of the kind depicted in Fig.\ref{fig:polrot} and consider what symmetry restrictions are consequently imposed on the linear scatterer. Since our demand implies that all the polarisation transformation matrices between input and output plane waves are diagonal regardless of the input and output directions, the conclusion is that, upon scattering, the system preserves the circular polarisation handedness of any plane wave. As it is shown in Sec. IV A of Ref.\cite{FerCor2012b}, the circular polarisation handedness of all the plane waves composing an electromagnetic field is one possible definition of a fundamental property of the field: electromagnetic helicity. Only when all the plane waves composing a field have the same handedness with respect to their momentum vector is the helicity of the field well defined, and can take the values $\pm 1$. As an operator, helicity is defined as the projection of the angular momentum onto the direction of the linear momentum \cite{Tung1985}, i.e. $\Lambda={\mathbf{J}}\cdot{\mathbf{P}}/|{\mathbf{P}}|$. For a single plane wave of momentum vector $\p$, the two possible states of definite helicity coincide with the two possible states of definite angular momentum along $\p$ ($J_{\p}$) with eigenvalues equal to $\pm 1$, and they also coincide with the two possible states of circular polarization handedness.

Our demands on the polarisation transformation properties of the scatterer have resulted in the scatterer being restricted to meet a conservation law: helicity conservation. Consequently, the electromagnetic response of the scatterer must be invariant under the transformation generated by the helicity operator: Electromagnetic duality \cite{Calkin1965,Zwanziger1968}.

In the same way that one of the components of angular momentum generates rotations along the corresponding axis\cite{Rose1957}, helicity is the generator of duality transformations:
\begin{equation}
\label{eq:DualityGeneralization}
\mathbf{E'}= \cos\gamma \mathbf{E} - \sin\gamma\mathbf{H},\
\mathbf{H'}= \sin\gamma \mathbf{E} + \cos\gamma\mathbf{H},
\end{equation}
where $\gamma$ is a real angle and the vacuum electric and magnetic constants are assumed to be equal to one. The typical exchange $\mathbf{E}\rightarrow\mathbf{H},\mathbf{H}\rightarrow-\mathbf{E}$ is recovered for $\gamma=-\pi/2$.
 Helicity preservation and invariance under the transformation in (\ref{eq:DualityGeneralization}) are hence equivalent conditions in the same way that angular momentum preservation along an axis is equivalent to rotational invariance along that same axis. The GRLP condition on all scattering directions imposed on our system has lead us to conclude that its electromagnetic response must be invariant under duality transformations.  

We would like to mention that a symmetry of a system and its associated conservation law imply preservation of the eigenstates of the generator of the symmetry, as explained in Sec. 4.1 of Ref \cite{Sakurai2006}. The different eigenstates of the generator do not mix after interaction with the system but, since they can pick up a different complex scaling during such interaction, a conservation law does not imply that the average value of the property represented by the generator of the symmetry remains unchanged.

To proceed with the study of the symmetries of our system, we introduce a concise notation that expresses the action of the system $S$ as a linear operator which takes input plane waves of a given momentum $\p$ and helicity $\pm$, $|\p,\pm\rangle$ into output plane waves $|\pbar,\pm\rangle$. 
\begin{equation}
\label{eq:decomp2}
S=\fourint \ \left(\aaa|\pbar,+\rangle\langle+,\p|+\ddd|\pbar,-\rangle\langle-,\p|\right).
\end{equation}
The orthogonality relationships $\langle \lambda,\p|\p',\lambda'\rangle=\delta_{\lambda-\lambda'}\delta(\p-\p')$ are assumed. The absence of helicity flipping cross-terms $|\pbar,-\rangle\langle+,\p|,|\pbar,+\rangle\langle-,\p|$ in (\ref{eq:decomp2}) reflects the helicity preservation condition. Taking a pair $(\p,\pbar)$, the $2\times 2$ sub-scattering matrix that specifies the conversion is
\begin{equation}\nonumber
\begin{bmatrix}\aaa& 0\\0&\ddd \end{bmatrix}
=\begin{bmatrix}|\aaa|\exp(i\arg\aaa) & 0\\0&|\ddd|\exp(i\arg\ddd) \end{bmatrix}
.
\end{equation}
The dependence of the $\aaa$ and $\ddd$ on the momenta $(\p,\pbar)$ allows for different transformations of the required diagonal kind with different angles of rotation $\beta_{\p}^{\pbar}={\left(\arg\aaa-\arg\ddd\right)}$.

We now ask the following question. Let us assume that there exists a pair $(\p,\pbar)$ for which $\beta_{\p}^{\pbar}\neq 0$. What can be said about the symmetries of the system? To answer this question, we consider the mirror operation $M_{\p\pbar}$ across the plane defined by $(\p,\pbar)$ and assume that the system possesses this mirror symmetry, i.e. it is invariant under the action of the mirror operator: $M_{\p\pbar}^{-1}SM_{\p\pbar}=S$. This particular mirror operation leaves the momentum vectors invariant because they are contained in the mirror plane and, since any spatial inversion operation flips the helicity value \cite{Tung1985}, the states transform as $M_{\p\pbar}|\p,\pm\rangle=|\p,\mp\rangle,M_{\p\pbar}|\pbar,\pm\rangle=|\pbar,\mp\rangle$. Using these transformation properties, and the fact that the mirror operator is unitary: $M_{\p\pbar}^{-1}=M_{\p\pbar}^{\dagger}$ we can see that, if the system is invariant under this mirror transformation, the angle of rotation $\beta_{\p}^{\pbar}$ is equal to zero.
\begin{equation}\nonumber
	\begin{split}
		\aaa&=\langle +,\pbar|S|\p,+\rangle=\langle +,\pbar|M_{\p\pbar}^{\dagger}SM_{\p\pbar}|\p,+\rangle\\&=\langle -,\pbar|S|\p,-\rangle=\ddd \Rightarrow \beta_{\p}^{\pbar}=0.
	\end{split}
\end{equation}

Therefore, if there exists a pair $(\p,\pbar)$ for which $\beta_{\p}^{\pbar}\neq 0$,  what we can say about the system is that it does not have that particular mirror symmetry: $M_{\p\pbar}^{-1}SM_{\p\pbar}\neq S$. Summarizing the discussion up to this point: By imposing that the system performs GRLP for any pair of input output plane waves and that the rotation is non zero for at least one pair, we have found that it must preserve helicity and lack of at least one mirror plane of symmetry. These are two necessary conditions for GRLP. Expressed with the help of commutators \bibnote{We recall here that the commutator between two operators $A$ and $B$, is defined as $[A,B]=AB-BA$, and that for unitary and hermitian operators $[A,B]=0$ if and only if $A^\dagger B A =B$.} between operators and denoting as $M_{\hat{\alpha}}$ the mirror operation across the plane perpendicular to vector $\hat{\alpha}$, our findings are:
\begin{equation}
\label{eq:nec}
\textrm{GRLP}\Longrightarrow [S,\Lambda]=0 \textrm{ and } \exists \ \hat{\alpha} \textrm{ s.t. } [S,M_{\hat{\alpha}}]\neq 0.
\end{equation}
It can be shown that the two necessary conditions are not sufficient in general. It can also be shown that, when further assuming rotational symmetry, the three conditions together are indeed sufficient for nonzero GRLP.

Having derived (\ref{eq:nec}) for a general linear system, we now turn to the study of GRLP by a mixture containing a large number of randomly oriented scattering particles immersed in an isotropic and homogeneous medium. We assume that the mixture has a linear response and contains only one kind of particle. Our aim is to investigate what conditions (\ref{eq:nec}) imposes on the scattering operator $S_u$ of the individual particle. For this, we make use of the theory of independent random scattering to compute the Mueller matrix of the mixture. The Mueller matrix relates the input Stokes parameters with the output Stokes parameters. The theory of independent random scattering \cite{Hulst1957,Tsang2004} is typically used to approximately describe electromagnetic propagation in a random solution of small scattering particles. It is strictly applicable when the individual particles are sufficiently separated\bibnote{In \cite[expr. 3.1.13]{Tsang2004} we find a condition on the standard deviation of the random distance $d_{ij}$ between two particles: $S.D.(d_{ij})\ge \frac{\lambda}{4}$. In \cite[chap. 1.21]{Hulst1957}, the condition for applying independent scattering is given in terms of the radius of the particles $R$:  $d_{ij}>>3R$.} and the number of particles tends to infinity. In this case, the Mueller matrix of the total solution $L_S(\p,\pbar)$ can be computed as the average sum of the Mueller matrices for all possible orientations of the individual particle. If $f(\cdot)$ is the function that converts a 2x2 scattering matrix to its corresponding Mueller matrix \cite{Fujiwara2007}, we have that
\begin{equation}
\begin{split}
\label{eq:L}
L_S(\p,\pbar)&=n_0 \int dR f(S_u^R(\p,\pbar))\\
	&=n_0 \int dR f(\langle\bar{\lambda},\pbar|R^{\dagger} S_u R|\p,\lambda\rangle).
\end{split}
\end{equation}
where $n_0$ is the density of particles per unit volume, $\int dR$ indicates the sum over all possible rotations and $S_u^R(\p,\pbar)$ is the 2x2 scattering matrix of a $R$-rotated version of the individual particle with coefficients $\langle\bar{\lambda},\pbar|R^{\dagger}S_uR|\p,\lambda\rangle$. It is important to note that due to the integral over all rotations, equation (\ref{eq:L}) is only exact in the limit of infinite number of particles. From now on, we will take (\ref{eq:L}) as an effective response for the mixture and comment on which of the obtained results explicitly rely on the $\int dR$ average and which do not.

We start with the condition concerning mirror symmetry: $\exists \ \hat{\alpha} \textrm{ such that } [S,M_{\hat{\alpha}}]\neq 0$. The Mueller matrix of the mirror system  can be shown to be
{\small
\begin{equation}
\label{eq:LM}
L_{M_{\hat{\alpha}}^\dagger SM_{\hat{\alpha}}}(\p,\pbar)=n_0 \int dR f(\langle\bar{\lambda},\pbar|R^{\dagger}M_{\hat{\alpha}}^\dagger S_uM_{\hat{\alpha}}R|\p,\lambda\rangle).
\end{equation}
} 
Lack of the mirror plane of symmetry $M_{\hat{\alpha}}$ for the mixture implies that $L_S(\p,\pbar)\neq L_{M_{\hat{\alpha}}^\dagger SM_{\hat{\alpha}}}(\p,\pbar)$ for at least one pair $(\p,\pbar)$. 

Now, let us assume that the individual particle possesses a symmetry of the rotation reflection kind \cite{Bishop1993}: $M_{\hat{\beta}}R_{\hat{\beta}}\left(\frac{2\pi}{m}\right)$, with $m$ a positive integer. These families of operators contain the common spatial inversion operations: For different values of $m$ we obtain parity, mirror symmetries and improper axes of rotation. When we assume any of these symmetries for $S_u$, the argument of $f(\cdot)$ in (\ref{eq:LM}) can be written\bibnote{To obtain such result,  write $M_{\hat{\alpha}/\hat{\beta}}$ as $M_{\hat{\alpha}/\hat{\beta}}=\Pi R_{\hat{\alpha}/\hat{\beta}}(\pi)$, substitute $S_u=R_{\hat{\beta}}^{\dagger}\left(\frac{2\pi}{m}\right) R_{\hat{\beta}}^{\dagger}(\pi)\Pi^{\dagger}S_u \Pi R_{\hat{\beta}}(\pi)R_{\hat{\beta}}\left(\frac{2\pi}{m}\right)$, use the crucial facts that the parity operator $\Pi$ commutes with any rotation and $\Pi^2$ is the identity, and group all fixed rotations into rotation $\tilde{R}$.} as $\langle\bar{\lambda},\pbar|R^{\dagger}\tilde{R}^\dagger S_u \tilde{R} R|\p,\lambda\rangle$, where $\tilde{R}$ is a fixed rotation. Then:
{\begin{equation}\nonumber
	\begin{split}
		L_{M_{\hat{\alpha}}^\dagger SM_{\hat{\alpha}}}(\p,\pbar)&=n_0 \int dR f(\langle\bar{\lambda},\pbar|R^{\dagger}\tilde{R}^\dagger S_u \tilde{R} R|\p,\lambda\rangle)\\
&=L_S(\p,\pbar),\ \forall \ (\p,\pbar).
	\end{split}
\end{equation}
}
\noindent The second equality follows from the fact that, when $R$ covers all possible rotations once, $\tilde{R}R$ also covers all possible rotations once and the result of the integral is always the same, independently of $\tilde{R}$ (including the case of the identity $\tilde{R}=I$). This is an application of the re-arrangement lemma from group theory \cite{Tung1985}.

We have just proved that for the mixture to lack one mirror symmetry $M_{\hat{\alpha}}$, the individual particle must not have any symmetry of the type $M_{\hat{\beta}} R_{\hat{\beta}}\left(\frac{2\pi}{m}\right)$. The lack of all of these symmetries (for all $m$) is equivalent to the particle being chiral \cite{Bishop1993}. Some qualitative consideration suffices to realize that such condition on the particle is also sufficient for the mixture as a whole to also lack all of the $M_{\hat{\beta}}R_{\hat{\beta}}\left(\frac{2\pi}{m}\right)$ symmetries and therefore become chiral. It is actually impossible for the random mixture to possess one mirror plane of symmetry without possessing them all. Since this result needs the averaging over random orientations, it will not apply to an ordered system. For instance, an ensemble of oriented molecules can easily lack one mirror plane of symmetry without lacking them all. 

\begin{figure}[h]
\subfloat{\includegraphics[scale=0.9]{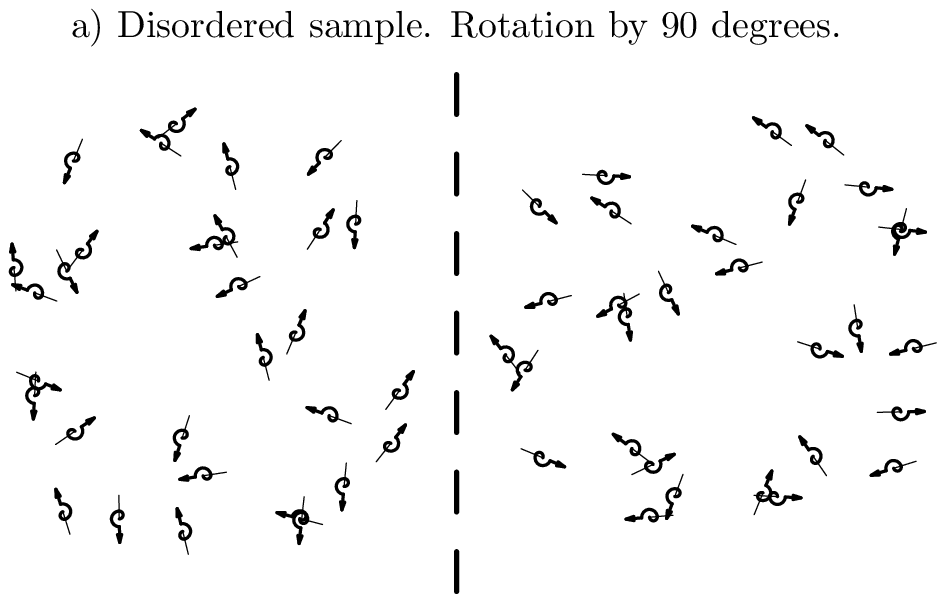}}
\vspace{0.75cm}
\subfloat{\includegraphics[scale=0.9]{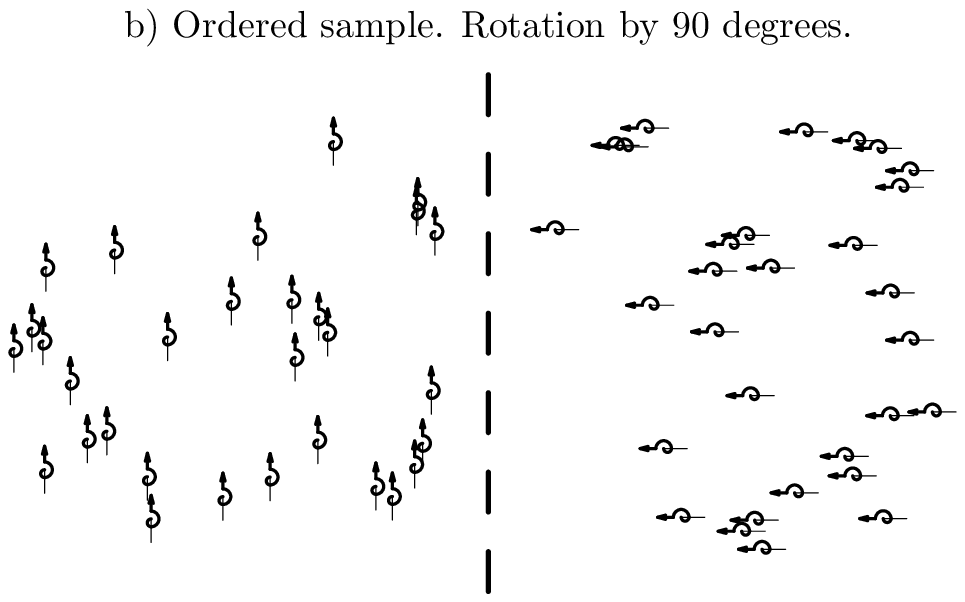}}
\caption{Illustration of the effective rotational symmetry acquired by a random solution. The left parts of the figures are the initial mixtures and the right parts are the rotated versions of the initial mixtures. Even though the requirement of infinite number of particles cannot be graphically illustrated, it can already be perceived for the finite number of particles in the figures that, after a rotation, the light scattering properties of an ordered sample (b) change dramatically, while those of a disordered sample (a) do not; in fact, under the assumptions made in the derivation contained in the text, they do not change at all when the number of particles tends to infinity. Note that the effective rotational symmetry in (a) is acquired independently of any property of the individual particle.}
\label{fig:helices}
\end{figure}

We now turn to the duality condition $[S,\Lambda]=0$. If we impose $[S_u,\Lambda]=0$, it is easy to show that $L_S(\p,\pbar)=L_{\Lambda^\dagger S\Lambda}(\p,\pbar)$ for all $(\p,\pbar)$. If all individual scatterers preserve helicity, clearly the overall response of the mixture will preserve helicity. Importantly, the $\int dR$ averaging does not need to be invoked in such proof. This result also applies to an ordered mixture or a mixture with a small number of particles. We now ask: Could it be that due to the averaging and/or randomness of the mixture, $[S_u,\Lambda]\neq0$ but $[S,\Lambda]=0$? We later prove that, in general, the answer is no, and that helicity preservation by the particle is necessary for helicity preservation by the mixture. A statement like $[S,\Lambda]=0$, containing the system scattering operator $S$ involves all input and output plane wave directions (see (\ref{eq:decomp2})), and implies the preservation of helicity in all directions. Since we have seen that helicity preservation is a necessary condition for GRLP and not all scatterers preserve helicity \cite{FerCor2012,FerCor2013}, these results seem at odds with the current understanding of molecular optical activity where chirality of the molecule is seen as the only necessary and sufficient condition. This conflict is resolved. We now show that the forward scattering direction is a special case since helicity is preserved independently of any property of the individual particle. Using again the re-arrangement lemma (which implies the $\int dR$ averaging) allows one to show that $L_S(\p,\pbar)=L_{\tilde{R}^\dagger S\tilde{R}}(\p,\pbar)$, for any rotation $\tilde{R}$. Such effective rotational symmetry implies conservation of the angular momentum along any axis and ensures helicity preservation in the forward scattering direction ($\p/|\p|=\pbar/|\pbar|$): Helicity being the angular momentum along the momentum axis of the input and output plane waves ($\Lambda=\mathbf{J}\cdot\mathbf{P}$), it must be preserved by a rotationally symmetric system when the two momenta share the same axis ($\p/|\p|=\pbar/|\pbar|$). Therefore, in the forward scattering direction, a solution of a chiral molecule can exhibit GRLP. Having used the average over all possible rotations $\int dR$ to derive this result, it will not apply to ordered systems or systems with a small number of particles. For example, the result does not apply to an ensemble of oriented molecules. The acquisition of effective rotational symmetry due to orientation randomness and its breaking by an ordered sample is illustrated in Fig. \ref{fig:helices}. In general, unless the individual molecules preserve helicity, the ensemble will not meet one of the necessary conditions for GRLP in an arbitrary scattering direction. As a consequence, those directions cannot exhibit polarisation rotations of the type depicted in Fig.\ref{fig:polrot}. These considerations match the results in \cite{Papakostas2003} and \cite{Ren2012}, where the interaction of light with an array of ordered nanostructures results in a polarisation transformation where, when interpreted as a rotation, the angle of rotation depends on the input polarization angle.

We use the Mueller matrix formalism to study random mixtures of different kinds of particles and provide analytical and numerical evidence that helicity is only preserved for all $(\p,\pbar)$ when the individual particles preserve helicity. For a helicity preserving system, the two Stokes vectors of well defined helicity $\begin{bmatrix} 1&0&0&\pm1\end{bmatrix}$ must be eigenvectors of the Mueller matrix $L$. This imposes restrictions to the matrix coefficients $L_{ij}$ which can be used to determined whether a particular Mueller matrix preserves helicity:

{\small
\begin{equation}
\label{eq:helpres}
L_{11}=L_{44},\ L_{14}=L_{41},\ L_{21}=L_{31}=L_{24}=L_{34}=0.
\end{equation}
}
In general, equation (\ref{eq:L}) must be evaluated numerically. We later provide numerically obtained values for mixtures of spherical, conical and helical particles. For the simple case of small (w.r.t. the wavelenght) spherical particles with relative electric constant $\epsilon$ and relative magnetic constant $\mu=1$, there is an analytical expression for $S_u(\p,\pbar)$ \cite{Tsang2004} which allows to compute $L_S(\p,\pbar)$,
{\footnotesize
\begin{equation}
\label{eq:sphere}
\begin{split}
&L_S(\p,\pbar)=\delta_{\p\pbar}I_{4\text{x}4}+n_0 k^2a^3\frac{\epsilon -1}{4\pi(\epsilon+2)}\times\\
&\times\begin{bmatrix}
\cos^2(\dtheta)+1 & \cos^2(\dtheta)-1 & 0 & 0\\
\cos^2(\dtheta)-1 & \cos^2(\dtheta)+1 & 0 & 0\\
0&0& 2\cos(\dtheta)&0\\
0&0&0& 2\cos(\dtheta)\end{bmatrix},
\end{split}
\end{equation}
}
\noindent where $k$ is the wavenumber, $a$ the radius of the sphere and $\dtheta$ is the angle between the input and output momentum vectors. The first term is the $4\times 4$ identity matrix, which, as indicated by the kronecker delta $\delta_{\p\pbar}$ is only added when $\p=\pbar$. It represents the contribution of the original input plane wave.

For general $\dtheta$, matrix (\ref{eq:sphere}) violates the helicity preserving conditions (\ref{eq:helpres}). Therefore, in general, a solution of small spheres does not preserve helicity. The breaking of duality symmetry can be traced back to the individual particle.

From a recently obtained result \cite{FerCor2012} regarding the conditions for duality symmetry (helicity preservation) on the macroscopic Maxwell's equations, we know that a particle of arbitrary shape with electric constant $\epsilon$ and magnetic constant $\mu$ immersed in a solvent $(\epsilon_s,\mu_s)$ would preserve helicity if and only if $\epsilon/\mu=\epsilon_s/\mu_s$. Then, particles with $\epsilon=2.25,\ \mu=1$ are non-dual when immersed in vacuum and hence do not preserve the helicity of light.

For the case of the small spheres in equation (\ref{eq:sphere}), the fact that helicity is not preserved by the individual particle makes the whole random mixture non helicity preserving. The randomness of the mixture does not help in terms of helicity preservation, except, as already explained, in the forward scattering direction. We conclude that a necessary condition for a solution of small spheres to preserve helicity is that the individual sphere preserves helicity. We have already discussed that such condition is also sufficient. 

To investigate whether the conclusions reached for small spheres also hold for mixtures of other kinds of particles and sizes, we numerically computed the rotational average (\ref{eq:L}) for small conical, and helical particles and for spheres of different sizes, with $\epsilon=2.25$ and $\mu=1$ immersed in vacuum.

To measure the degree of helicity transformation in each case we use the following metric on the resulting Mueller matrices:
{\footnotesize
\begin{equation}
\label{eq:gamma}
\begin{split}
\Gamma = & \frac{\left(L_{11}+L_{14}-\left(L_{41}+L_{44}\right)\right)^2}{2\left(L_{11}+L_{14}-\left(L_{41}+L_{44}\right)\right)^2+\left(L_{11}+L_{14}+\left(L_{41}+L_{44}\right)\right)^2}+\\
&\frac{\left(L_{11}-L_{14}+\left(L_{41}-L_{44}\right)\right)^2}{2\left(L_{11}-L_{14}+\left(L_{41}-L_{44}\right)\right)^2+\left(L_{11}-L_{14}-\left(L_{41}-L_{44}\right)\right)^2}.
\end{split}
\end{equation}
}
The first (second) line in (\ref{eq:gamma}) is the relative helicity change effected by the Mueller matrix on a Stokes vector of well defined positive (negative) helicity. $\Gamma=0$ for a helicity preserving Mueller matrix (\ref{eq:helpres}), and $\Gamma=1$ for a helicity flipping Mueller matrix.

\begin{figure}[h]
\subfloat[][]{\includegraphics[scale=1]{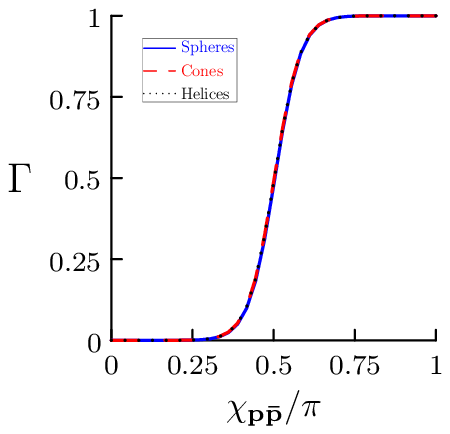}}
\subfloat[][]{\includegraphics[scale=1]{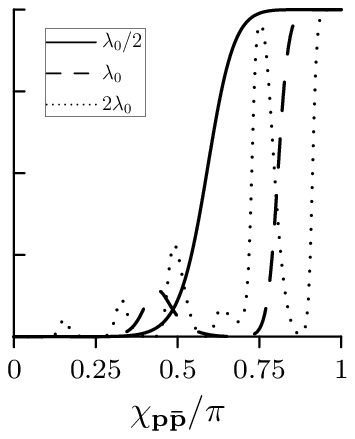}}
\caption{(Color online) Helicity transformation metric $\Gamma(\dtheta)$ results obtained from the numerical computation of the Mueller matrices of vacuum solutions of different kinds of particles with $\epsilon=2.25$ and $\mu=1$. $\Gamma=0$ corresponds to helicity preservation and $\Gamma=1$ to complete helicity change. $\Gamma=0$ for all $\dtheta$ would correspond to a helicity preserving solution. (a) Spheres, cones and helices of dimension $\approx \lambda_0/6$. The wavelength of light was $\lambda_0=632$ nm. (b) Spheres with diameters $\approx [\lambda_0/2,\lambda_0,2\lambda_0]$. All cases break helicity preservation. In the small particle case (a), the results are very similar (small differences not visible in the figure) independently of the kind of particle. This can be explained because all small particles can be treated in the dipolar approximation. When the size of the particles is comparable to the wavelength (b), we observe a more complex behavior of $\Gamma(\dtheta)$ with oscillations where $\Gamma$ is close to zero for some angles. We speculate that this behavior is related to the excitation of higher (than dipole) multipolar moments. }
\label{fig:final}
\end{figure}

Figure \ref{fig:final}(a) plots $\Gamma$ as a function of the relative angle between the input and output momenta ($\dtheta$) for spheres, cones and helices of dimensions $\approx \lambda_0/6$, where $\lambda_0$ is the wavelength\bibnote{The individual scattering matrices $S^R_u(\p,\pbar)$ for the computation of (\ref{eq:L}) were obtained by illuminating the single scatterer with plane waves of defined helicity from all input directions described by polar and azimuthal angles $(\theta,\phi)$. For each input direction, the scattered far field in all output directions $(\bar{\theta},\bar{\phi})$ was then computed. For the sphere, the computation was done analytically by applying the Mie scattering theory. For the cylinder and the helix, a commercial finite elements package was used in which far field calculations are made with the Stratton-Chu formula. Given an input output momenta pair $(\theta,\phi),(\bar{\theta},\bar{\phi})$, the helicity scattering coefficients are asymptotically proportional to the projection of the far field for direction $(\bar{\theta},\bar{\phi})$ onto the corresponding helicity vector basis. For the sphere the integral $\int dR$ in (\ref{eq:L}) is trivial. For the cone and the helix, it must be numerically computed. All angles were discretized using 5 degrees intervals.}. Fig. \ref{fig:final}(b) plots $\Gamma(\dtheta)$ for spheres of diameters $\approx [\lambda_0/2,\lambda_0,2\lambda_0]$. All cases show that the solutions do not preserve helicity, strongly suggesting that, in general, for the solution to preserve helicity, the individual particles must preserve helicity. Note how helicity preservation properties are independent of the geometrical properties of the particles. Spheres have all mirror planes of symmetry, helices lack them all and cones have some but not all. In all cases $\Gamma(0)=0$ in full agreement with our previous discussion about forward scattering ($\p/|\p|=\pbar/|\pbar|$). In all cases $\Gamma(\pi)=1$, indicating that for backward scattering ($\p/|\p|=-\pbar/|\pbar|$), helicity is always exactly flipped. This is actually a general result due to the effective rotational symmetry of the solutions. The angular momentum along $-\pbar$ must be the same as the input angular momentum along $\p$ since they share the same axis. Since helicity is $\Lambda=\mathbf{J}\cdot\mathbf{P}/|\mathbf{P}|$, helicity must exactly change sign due to the preservation of $\mathbf{J}$ and the sign change in $\mathbf{P}$ in the backward scattering direction $\p/|\p|=-\pbar/|\pbar|$.

The current understanding of the phenomena of molecular optical activity, when the molecules are in a disordered solution, is that chirality of the individual molecule is the only necessary and sufficient condition \cite{Oloane1980,Bishop1993}. Helicity preservation (duality symmetry) is not given a role. This apparent conflict is completely resolved: As we have discussed, the large number of randomly oriented particles endows the mixture with an effective rotational symmetry, which is shown to lead to the conservation of helicity in the forward scattering direction independently of any property of the individual particle. Therefore, in the forward scattering direction, a solution of a chiral molecule can exhibit optical activity without the individual molecule having to preserve helicity. In an ordered system or in any non-forward scattering direction the effective rotational symmetry disappears and explicit helicity preservation (electromagnetic duality symmetry) by (of) the individual particle in the solvent is needed for optical activity in the sense used in this article. 

In his seminal work \cite{Condon1937}, Condon posed a, to the best of our knowledge, still unresolved question: ``The generality of the symmetry argument is also its weakness. It tells us that two molecules related as mirror images will have equal and opposite rotatory powers, but it does not give us the slightest clue as to what structural feature of the molecule is responsible for the activity. Any pseudoscalar associated with the structure might be responsible for the activity and the symmetry argument would be unable to distinguish between them.''. Our answer to Condon's question is that helicity is the sought after pseudoscalar and that, when considering a single molecule (not a solution), there are two structural features that are necessary for the single molecule to be optically active in the sense used in this article: electromagnetic duality symmetry and lack of at least one mirror symmetry. If the molecule can be modeled as a dipolar scatterer, the electromagnetic duality symmetry condition restricts its polarisability tensor \cite{FerCor2013}. According to the results of this paper, these conditions also apply to the polarisability tensors of the individual inclusions in structured arrays designed to achieve artificial optical activity.

In this article, we have used the formalism of symmetries and conserved quantities to study a class of electromagnetic transformations which we have named generalized rotation of linear polarization (GRLP). We have identified two symmetry conditions necessary for an otherwise general electromagnetic linear scattering system to exhibit GRLP in all scattering directions: Lack of at least one mirror plane of symmetry and duality symmetry (helicity preservation). For the case of a random mixture of a single species of particle immersed in an isotropic and homogeneous medium, we have investigated the restrictions that the two necessary conditions impose on the individual scattering properties of the particle. We have proved that the individual particle must be chiral. We have also seen that helicity preservation in the forward scattering direction is provided by the randomness of the mixture independently of the properties of the individual particle. On the other hand, we have shown that for helicity preservation in a general scattering direction the individual particle in the solvent must itself have an electromagnetic response invariant under duality transformations, that is, it must preserve the helicity of light. Our results advance the current understanding of the phenomenon of molecular optical activity. Additionally, the general conditions in (\ref{eq:nec}) together with the results of \cite{FerCor2012} and  \cite{FerCor2013} provide insight that may assist in the design of polarization control devices, particularly at the nanoscale where metamaterials are used to engineer effective electric and magnetic constants.

\providecommand*\mcitethebibliography{\thebibliography}
\csname @ifundefined\endcsname{endmcitethebibliography}
  {\let\endmcitethebibliography\endthebibliography}{}

{\bf Acknowledgements}
This work was funded by the Australian Research Council Discovery Project DP110103697 and the Center of Excellence for Engineered Quantum Systems (EQuS). G.M.-T is also funded by the Future Fellowship program (FF).
\end{document}